\documentclass[12pt,fleqn]{article}
\usepackage{ifthen}                   
\usepackage{exscale}                  
\usepackage[intlimits]{amsmath}       
\usepackage{amsfonts}
\usepackage{amssymb,amscd}
\usepackage[dvips]{epsfig}                   
\usepackage{array}
\usepackage{wrapfig}

\def\Quadrat#1#2{{\vcenter{\hrule height #2
  \hbox{\vrule width #2 height #1 \kern#1
    \vrule width #2}
  \hrule height #2}}}
\def\dAlemb{\mathop{\kern 1pt\hbox{$\Quadrat{8pt}{0.4pt}$} \kern1pt}}
\def\dAlember{\mathop{\kern 1pt\hbox{$\Quadrat{4pt}{0.4pt}$} \kern1pt}}
\def\Chi{{\mathop{\kern 2pt\vcenter{\hbox{$\chi $}}\kern2pt}}}
\def\dslash{\partial\kern-.5em\slash}
\def\kslash{k\kern-.5em\slash}
\def\slh#1{#1\kern-.5em\slash}
\def\fslash#1{#1 \kern-.5em\slash}

\def\fbar#1{#1\kern-.5em\raise6pt\hbox{\footnotesize /}}

\newcommand{\lsim}{\mathrel{\rlap{\lower4pt\hbox{\hskip1pt$\sim$}}
\raise1pt\hbox{$<$}}}

\newcommand{\nn}{\nonumber}

\newcommand{\SSc}{\scriptscriptstyle}
\newcommand{\be}{\begin{eqnarray}}
\newcommand{\ee}{\end{eqnarray}}

\newcommand{\Pslash}{P\hspace{-.5em}/\hspace{.15em}}
\newcommand{\pslash}{p\hspace{-.5em}/\hspace{.15em}}

\newcommand{\fourint}[1]{\int\!\frac{d^4 #1}{(2\pi)^4}}

\newcommand{\vect}[1]{{\mbox{\boldmath $#1$}}}

\begin{document}

\setlength{\baselineskip}{20pt}

\psfull

{\small

\rightline{UNITUE--THEP-01/00}
\rightline{hep-ph/0001261}}

\vspace{1cm}

\begin{center}
\begin{large}
{\bf A Comparison between Relativistic and Semi--Relativistic
Treatment in the Diquark--Quark Model$^\dagger$}\\
\end{large}
\vspace{1cm}
{\bf 
M.\ Oettel$^1$ and R.\ Alkofer$^2$
\\}
\vspace{0.2cm}
{Institute for Theoretical Physics, T\"ubingen University \\
Auf der Morgenstelle 14, D-72076 T\"ubingen, Germany
}
\end{center}
\vspace{1cm}
\begin{abstract}
\normalsize
\noindent
In the diquark--quark model of the nucleon including scalar and axialvector
diquarks we compare solutions of the ladder Bethe--Salpeter equation in the
instantaneous Salpeter approximation and in the fully covariant (i.e.\
four--dimensional) treatment.  We obtain that the binding energy is severely
underestimated in the Salpeter approximation. For the electromagnetic form 
factors of the nucleon we find that in both approaches the overall shapes 
of the respective form factors are reasonably similar
up to $Q^2 \approx 0.4$ GeV$^2$. However, the magnetic 
moments differ substantially as well as results for the pion--nucleon
and the axial coupling of the nucleon. 
\end{abstract}
\vspace{2.5cm}
\vfill

\noindent
\rule{5cm}{.15mm}

\noindent
$^\dagger$Supported by COSY under contract 41376610 and DFG 
under contract We 1254/4-1.
\\
$^1$E-Mail:oettel@pion06.tphys.physik.uni-tuebingen.de\\
$^2$E-Mail:reinhard.alkofer@uni-tuebingen.de

\newpage

\pagenumbering{arabic}


Bound states of two particles in quantum field theory can be described by 
solutions of integral equations such as the Bethe--Salpeter (BS) equation
\cite{Bet51}.
It has the generic form
\begin{equation}
 \Psi(p,P)= S_1(p_1)\, S_2(p_2) \fourint{p'} K(p,p',P) \Psi(p',P) \, ,
\end{equation}
where 
$S_1$ and $S_2$ denote the fully dressed propagators of particle 1 and 2, and
$K$ is the full, 2P--irreducible interaction kernel between the in- and
outgoing particles. The wave function (or Bethe--Salpeter amplitude) $\Psi$
is not an observable quantity but it can be employed to compute
e.g.\ form factors as matrix elements of operators \cite{Man55}. 

Little success has been made in going beyond the ladder approximation
(see, however, e.g.\ \cite{The99})
where the propagators $S_1$ and $S_2$ are approximated by the corresponding 
bare propagators
and the kernel $K$ just contains the interaction to lowest order which might
be the exchange of a third particle with bare propagator $S_3$:
\begin{equation}
 \Psi(p,P)= S_1^{bare}(p_1)\, S_2^{bare}(p_2) \fourint{p'} S_3(p,p',P)
  \Psi(p',P).
\end{equation}
Even in this approximation an analytical solution to the BS equation has been
found only for the Wick--Cutkosky model with a bound state of zero mass 
(for an introduction into this model see e.g.\ \cite{Sil98}).
 Moreover, practical attempts to solve the BS equation
in various models have resorted to further approximations, especially
three--dimensional reductions (for a critical comparison of different
three--dimensional reductions, see \cite{Bij97}).
The Salpeter approximation in which retardation effects in $S_3$ are neglected
reduces the BS equation to a Klein--Gordon or Dirac equation with a hermitian 
Hamiltonian. It is often claimed to be covariant though the neglection of 
the time--component of the four--momenta $p$ and $p'$
seems to prohibit this phrasing. It is true, however, that
the solutions obey relativistic translation invariance.

In this letter we choose a diquark--quark model of the nucleon for a 
comparison between the Salpeter approximation and the fully relativistic BS
solution. This has two reasons: If one assumes separable diquark correlations,
the three--body Faddeev equations reduce to a ladder BS equation which sums up
the quark exchange between the spectator quark and the diquark quasiparticle
\cite{Cah89,Rei90,Buc92,Ish95,Kus97,Oet98}.
Thus the truncation of the BS kernel to the one--particle exchange is not
completely arbitrary but directly linked to the separability assumption.
Secondly, a comparison between relativistic and semi--relativistic solutions in
the BS equations should not stop at the level of the solutions to the equations
but involve observable quantities. There are several attempts to calculate
observables within this framework, such as the electromagnetic form factors.
This has been done in the Salpeter
approximation \cite{Kei96a,Kei96b} and using the full solution
\cite{Hel97a,Oet99}. In the following we will compare the electromagnetic form
factors of the nucleon obtained in the Salpeter approximation to the ones
using the full BS equation employing model parameters as in refs.\
\cite{Kei96a,Kei96b}.

\goodbreak

\paragraph{The diquark-quark model\\}

The underlying approximations of the diquark--quark model are, first, the 
neglection of any 3--particle irreducible interactions between the quarks in
the nucleon, and, secondly, the separability of correlations in the two--quark
channel. The first assumption allows to derive a relativistic Faddeev equation
for the 6--point quark Green's function, and the second one reduces it to an 
effective quark--diquark BS equation.

The full derivation of the quark--diquark BS equation within the NJL model
can be found in refs.\ \cite{Rei90,Ish95}. In the following we will sketch the 
most important steps. As usual in the treatment of the BS equation we will work
in Euclidean space.
Scalar and axialvector diquarks are introduced via 
the assumption of a separable 4--point quark function:
\begin{eqnarray}
G_{\alpha\gamma , \beta\delta}^{\hbox{\tiny sep}}(p,q,P) \, :=
 \chi_{\gamma\alpha}(p) \,D(P)\,\bar \chi_{\beta\delta}(q) \; +\;
\chi_{\gamma\alpha}^\mu(p) \,D^{\mu\nu}(P)
    \bar  \chi_{\beta\delta}^\nu(q) \nn \; .   
\end{eqnarray}
$P$ is the total momentum of the incoming and the outgoing quark--quark pair,
$p$ and $q$ are the relative momenta between the quarks in the two channels.
$\chi_{\alpha\beta}(p)$ and $\chi_{\alpha\beta}^\mu(p)$  are vertex functions
of quarks with a scalar and an axialvector diquark, respectively. They belong
to a $\bar {\bf 3}$--representation in color space and are either flavor
antisymmetric (scalar diquark) or flavor symmetric (axialvector diquark). For
their Dirac structure we will retain the dominant contribution only. Thus we
introduce one scalar function $P(p)$ which depends only on the relative 
momentum $p$ between the quarks, 
\begin{eqnarray} 
\chi_{\alpha\beta}(p)&=&g_s
(\gamma^5 C)_{\alpha\beta}\; P(p) \; , \label{dqvertex_s} \\
\chi_{\alpha\beta}^\mu(p)&=&g_a (\gamma^\mu C)_{\alpha\beta}\; P(p).
\label{dqvertex_a} 
\end{eqnarray}
$C$ denotes hereby the charge conjugation
matrix, and $g_a$ and $g_s$ are effective coupling constants obtained by
normalization of the diquark states.
The Pauli principle requires the 
relative momentum to be defined $p=\frac{1}{2}(p_\alpha-p_\beta)$, 
where $p_\alpha$ and $p_\beta$ are the quark momenta \cite{Oet99}.
$P(p)$ parametrizes the extension of the vertex in momentum
space.    To facilitate the
comparison between our work and that of \cite{Kei96a,Kei96b} we choose
\begin{equation} 
P(p)= \exp ( -4\lambda^2p^2)\, . 
\end{equation} 
The propagators
of scalar and axialvector diquark are the ones for a free spin-0 and spin-1
particle, 
\begin{eqnarray} D(p)&=& -\frac{1}{p^2+m_{sc}^2} \, , \label{Ds}\\
D^{\mu\nu}(p)&=& -\frac{\delta^{\mu\nu}+ (1-\xi) \frac{p^\mu p^\nu}{m_{ax}^2} }
{p^2+m_{ax}^2} \, .  
\label{Da}  
\end{eqnarray} 
$\xi$ is a gauge parameter
introduced in \cite{LY62}, and in the following we will put $\xi=1$.
The constituent quark propagator is simply the free fermion propagator
\begin{equation} 
S(p)= \frac{ i\pslash -m_q}{p^2+m_q^2} \, . \label{S}      
\end{equation}

The nucleon BS wave function can be described by an effective multi--spinor 
characterizing the scalar and axialvector correlations (see e.g. \cite{Oet98}),
\begin{equation}
 \Psi (p,P) u (P)= 
    \begin{pmatrix} \Psi^5 (p,P) \\ \Psi^\mu  (p,P) \end{pmatrix} u(P)
\end{equation}
where $u(P)$ is a positive--energy Dirac spinor with $P$ being the total 
momentum of the bound state. $p$ is the relative momentum
between quark and diquark, respectively. The vertex function
is defined by amputating the legs off the wave function,
\begin{equation}
 \begin{pmatrix} \Phi^5  \\ \Phi^\mu \end{pmatrix} = 
    S^{-1}  \begin{pmatrix} D^{-1} & 0 \\ 0 & (D^{\mu\nu})^{-1} \end{pmatrix} 
 \begin{pmatrix} \Psi^5  \\ \Psi^\nu \end{pmatrix} . 
\end{equation}

The BS equation is a system of equations for wave and vertex function
that takes the compact form
\begin{equation}
  \fourint{p'} G(p,p',P) 
  \begin{pmatrix}\Psi^5 \\ \Psi^{\mu'}\end{pmatrix}(p',P) =0
  \label{bse_eq}
\end{equation}
where the object $G(p,p',P)$ involves the propagators of quark and diquark
and the interaction kernel that describes the quark exchange between quark and
diquark,
\begin{eqnarray}
 G (p,p',P) &=& (2\pi)^4 \delta(p-p') S^{-1}(p_q)
 \begin{pmatrix} D^{-1} & 0 \\ 0 & (D^{\mu\mu'})^{-1}\end{pmatrix} (p_d) 
 + \nonumber \\
 & & \frac{1}{2}
  \begin{pmatrix} \chi S^T(q)\bar\chi & -\sqrt{3} \chi^{\mu'} S^T(q)\bar\chi \\
    -\sqrt{3}\chi S^T(q)\bar\chi^{\mu} &  -\chi^{\mu'} S^T(q)\bar\chi^\mu
     \end{pmatrix}.
 \label{Gdef}
\end{eqnarray}

The partitioning of the momentum between quark and diquark introduces
a parameter $\eta$ with $p_q=\eta P+p$ and $p_d=(1-\eta)P - p$. Therefore
the momentum of the exchanged quark is $q=-p-p'+(1-2\eta)P$.
The relative momentum of the quarks at the diquark vertex $\chi$ is
$p_2=p+p'/2-(1-3\eta)P/2$ and the one at the conjugated vertex
$\bar\chi$ is $p_1=p/2+p'-(1-3\eta)P/2$.
 Relativistic translation
invariance requires that if $\Phi(p,P;\eta_1)$ is a solution of the
BS equation then also $\Phi(p+(\eta_2-\eta_1)P,P;\eta_2)$ is one.

The BS equation (\ref{bse_eq}) is solved in the rest frame of the bound state,
$P=\begin{pmatrix} \vect 0 \\ iM \end{pmatrix}$. In this frame the Salpeter
approximation amounts to neglecting the fourth component of all
vectors appearing in the interaction, {\it i.e} of the vectors $q$, $p_1$
and $p_2$. The immediate consequence is that the vertex function
will depend only on the relative three-momentum, $\Phi (p,P) \equiv
\Phi(\vect p)$.

\paragraph{Numerical Solutions\\}
In the following we use the complete partial wave decomposition of wave and 
vertex function for octet baryons given in \cite{Oet98},
\begin{eqnarray}
\left( \begin{array}{c}
\Phi^5(p,P) \\
\Phi^{4} (p,P) \\
{\vect{\Phi}} (p,P)\\ 
\end{array}\right)
=
\left( \begin{array}{c}
\left( \begin{array}{cc}
{\bf 1} \, {S}_1  & 0\\
\frac{1}{p}({\vect{\sigma}}\vect{p})\, {S}_2 & 0 \\ 
\end{array}\right)  \\
\left( \begin{array}{cc}
\frac{1}{p}(\vect{\sigma}\vect{p})\, {A}_1 & 0\\ 
{\bf 1} \, {A}_2 & 0\\
\end{array}\right) \\
\left( \begin{array}{cc}
i\hat{\vect{p}}(\vect{\sigma}\hat{\vect{p}}){A}_3 + 
   (\vect{\sigma} \times \hat{\vect{p}})
   (\vect{\sigma}\hat{\vect{p}}){A}_5 & 0\\
\frac{i}{p}\vect{p}{A}_4 + \frac{1}{p}(\vect{\sigma} \times 
{\vect{p}}) {A}_6 & 0 \\
\end{array}\right) 
\end{array}\right)
.
\label{expnuc}
\end{eqnarray} 
The unknown scalar functions $S_i$ and $A_i$ are functions of $p^2=p^\mu
p^\mu$ and of the angular variable $z= \hat P \cdot \hat p$ which denotes
the cosine of the angle between $p^\mu$ and the 4-axis. As explained above,
the functional dependence collapses in the Salpeter approximation from
the two variables ($p^2,z$) to the single variable $p^2(1-z^2)$. 
We expand the scalar functions in terms of Chebyshev polynomials of the 
first kind in the variable $z$,
\begin{equation}
 S_i [A_i] (p^2,z) = \sum_{n=0}^\infty i^n S_i^n [A_i^n] (p^2) T_n (z),
\end{equation}
and derive a system of coupled integral equations for
the Chebyshev momenta $S_i^n [A_i^n]$ that we solve iteratively as described 
in \cite{Oet98,
Oet99}. This expansion is very close to the hyperspherical expansion
that has been shown to work extraordinarily well in the massive Wick--Cutkosky
model \cite{Tjon96,Ahl99} and in quenched QED \cite{Ahl99}. 
This finding has been corroborated
by \cite{Oet98} in the diquark--quark model for pointlike diquarks.

As stated already we adopt the parameters of refs.\  \cite{Kei96a,Kei96b}.
In the case of the scalar--axialvector diquark model this especially includes
identical diquark normalizations leading to equal diquark--quark--quark
couplings in the scalar and axialvector channel, $g_s=g_a$. The BS equation 
(\ref{bse_eq}) can be written as an eigenvalue problem for $g_s$. 
This coupling is adjusted to yield the physical nucleon mass $M$=0.939 GeV
for given values of the quark and diquark mass as well as the diquark width
$\lambda$. 
In table \ref{par} we have listed the two sets of parameters, scalar diquark
only and scalar--axialvector diquark model, with their corresponding 
eigenvalues obtained by us in the full calculation and the Salpeter 
approximation. The values in parentheses are the ones from refs.\  
\cite{Kei96a,Kei96b}. Please note that due to a different
flavor normalization these values had to be multiplied by $\sqrt{2}$ to be
directly comparable to ours.

\begin{table}
\begin{tabular}{clccccl} \hline
Set & & $m_q$& $m_{sc}$& $m_{ax}$ & $\lambda$& $g_s$ \\
   & & ~~~[GeV]~~~ & ~~~[GeV]~~~ & ~~~[GeV]~~~ & ~~~[fm]~~~ & \\ \hline
I & Salpeter~~~~ & 0.35 & 0.65 & - & 0.18 & 20.0 ~~(20.0) \\
  & full     & & & & & 16.5 \\ \hline
II & Salpeter~~~~ & 0.35 & 0.65 & 0.65 & 0.24 & 12.3 ~~(11.5) \\
  & full     & & & & & \phantom{1}9.6 \\ \hline
\end{tabular}
\caption{The two parameter sets of the model.}
\label{par}
\end{table} 
 
Although we could reproduce the eigenvalue for the model case with scalar
diquark only, this is not the case for Set II. We observe that the
calculations of \cite{Kei96b} involved only 4 instead of 6 axialvector
components of $\Phi^\mu$, namely the projected ones onto
zero orbital angular momentum and the corresponding lower components.
Still one would expect a higher eigenvalue in the reduced system.
{\it The more striking observation is the amplification of the eigenvalue by 
about 20...25\% in the Salpeter approximation although the binding energy
is small,} being only 6\% of the sum of the constituent masses.
This is in contrast to results obtained in the massive Wick-Cutkosky model
\cite{Nieu96} where the Salpeter approximation leads to a reduction of the 
eigenvalue. This may be attributed to the exchange of a boson instead of
a fermion as in the present study.

\begin{figure}
 \begin{center}
  \epsfig{file=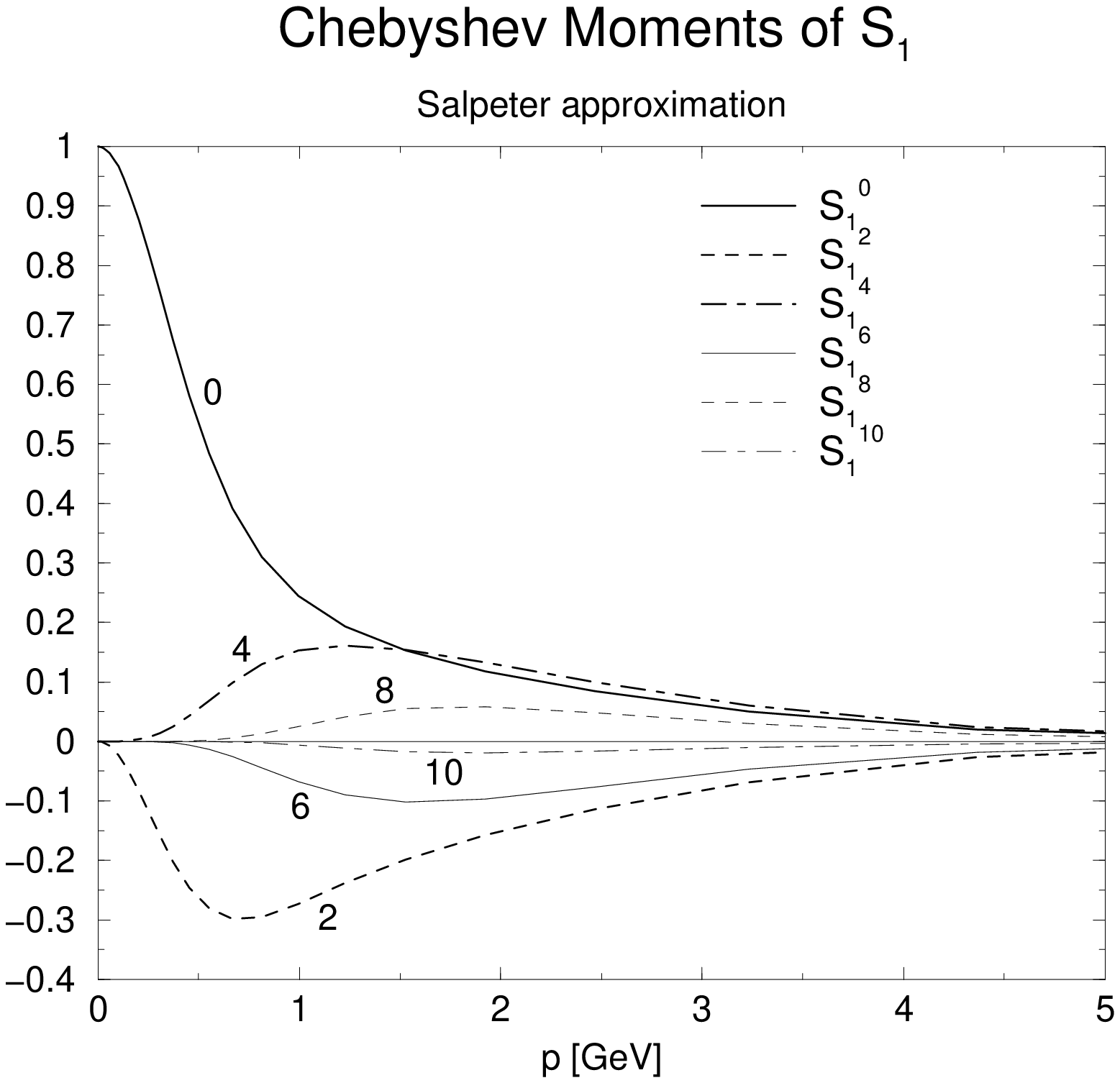,width=6cm}
  \epsfig{file=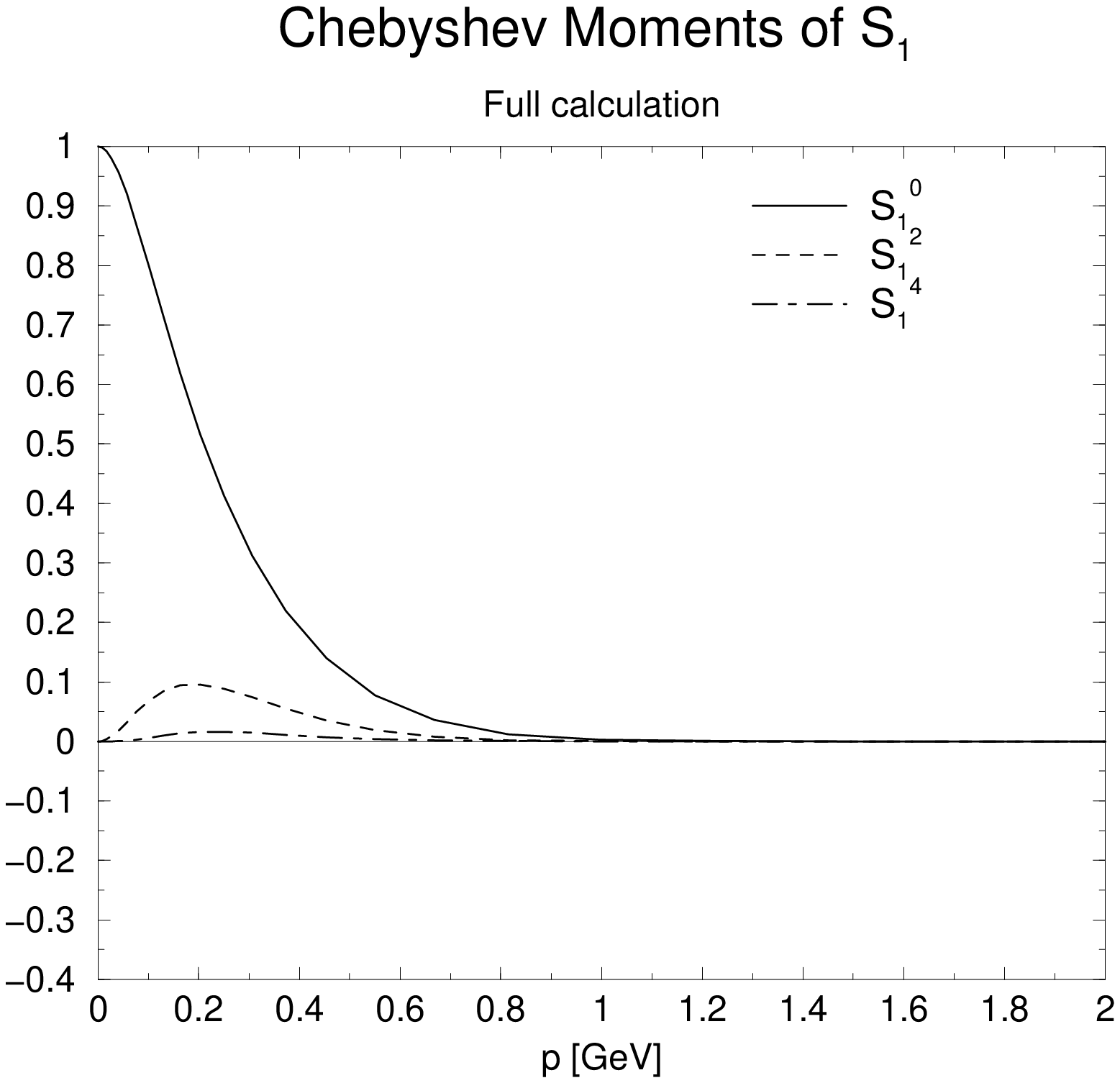,width=6cm}
 \end{center} 
 \caption{The Chebyshev expansion of the dominat scalar function $S_1(p,z)$
  in the Salpeter approximation and the full calculation.}
 \label{s1}
\end{figure}
 
The substantial difference between the two approaches is also reflected
in the vertex function solutions themselves. 
Fig. \ref{s1} shows the Chebyshev moments
of the dominant scalar function $S_1$ for both methods using the parameters
of Set I. Only the even momenta are given, since the odd ones are zero in the
Salpeter approximation (but are present, of course, in the full calculation).
Two things are manifest: The Salpeter amplitudes have a much broader spatial
extent than the full amplitudes. Secondly, the expansion in Chebyshev 
polynomials that relies on an approximate $O(4)$ symmetry converges much more
rapidly for the full solution but is hardly convincing in the Salpeter
approximation. Again, since the scalar functions depend in the Salpeter
approximation on just one variable, $p^2(1-z^2)$, our expansion is a 
cumbersome way of visualizing the solution but makes clear that the 
Salpeter approximation strongly violates the $O(4)$-symmetry.

\paragraph{Electromagnetic Form Factors\\}

The Sachs form factors $G_E$ and $G_M$ can be extracted from the solutions
of the BS equations using the relations
\begin{eqnarray}
 G_E =\frac{M}{2P^2} \mbox{Tr} \langle J^\mu \rangle P^\mu, \quad
 G_M =\frac{iM^2}{Q^2}  \mbox{Tr}\langle J^\mu \rangle \gamma^\mu_T, \\
 \langle J^\mu \rangle= \fourint{p_f} \fourint{p_i}
   \bar \Phi^T (P_f,p_f)  J^\mu  \Phi (P_i,p_i). \label{gdef}
\end{eqnarray}
with the definitions $P=(P_i+P_f)/2$ and $\gamma^\mu_T=\gamma^\mu- 
\hat P^\mu \hat \Pslash$. To this end, one has to normalize the wave and
the vertex function through the condition
\begin{equation}
  -\int \frac{d^4\,p}{(2\pi)^4}
  \int \frac{d^4\,p'}{(2\pi)^4}
   \bar \Psi(p',P_n) \left[ P^\mu \frac{\partial}{\partial P^\mu}
    G (p',p,P) \right]_{P=P_n} \Psi(p,P_n) \stackrel{!}{=} M \Lambda^+
\end{equation}
which uses the definition of $G$ given in eq. (\ref{Gdef})
and employs the positive--energy projector $\Lambda^+$.
This normalization integral is again performed easily in the
rest frame of the bound state, additionally the double integral over
the interaction kernel drops out in the Salpeter approximation since
the kernel is independent of $P$ in the rest frame. 

\begin{figure}
\begin{center}
 \epsfig{file=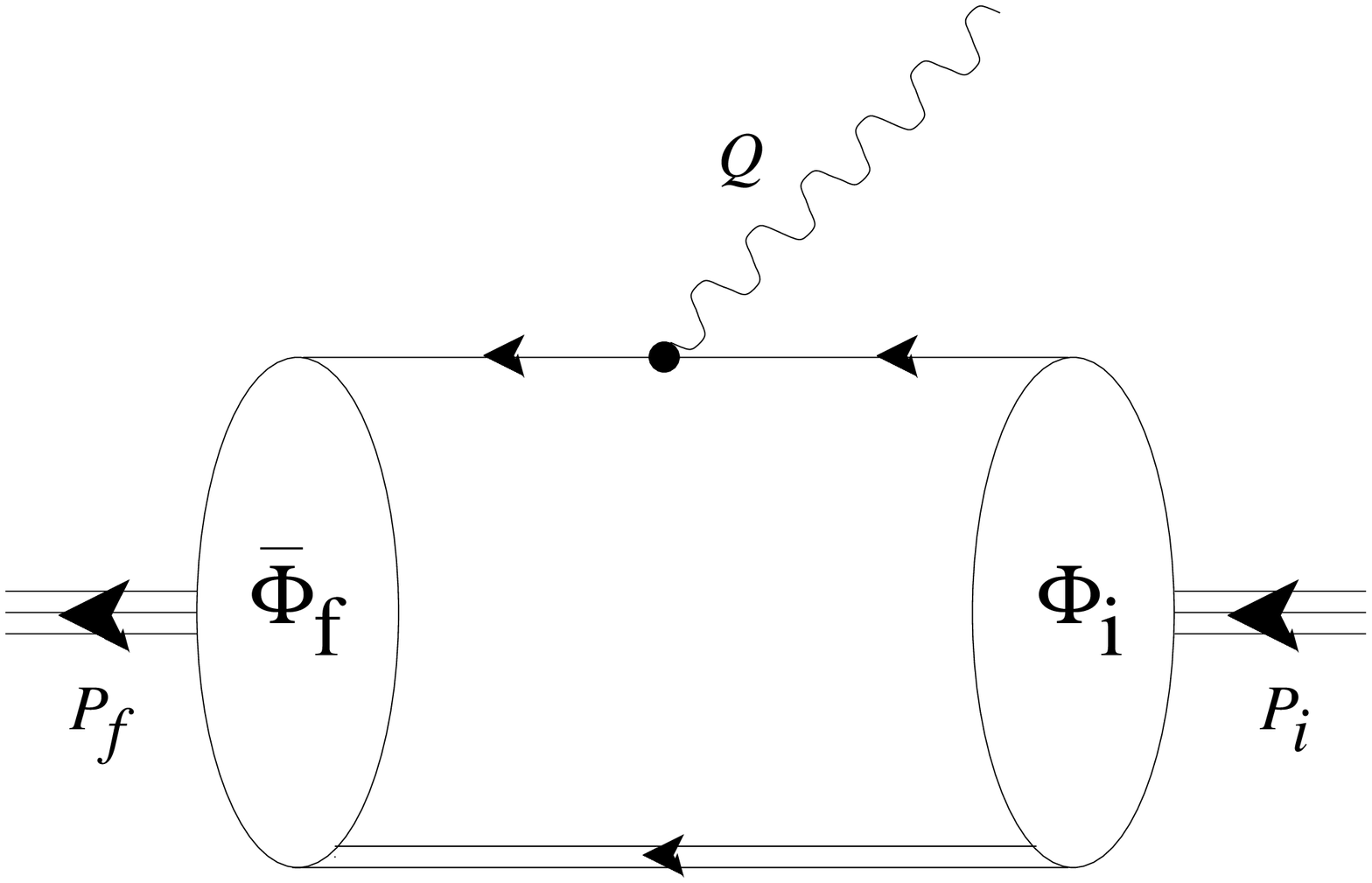,width=4cm}
 \begin{minipage}[t]{4cm}
  \epsfig{file=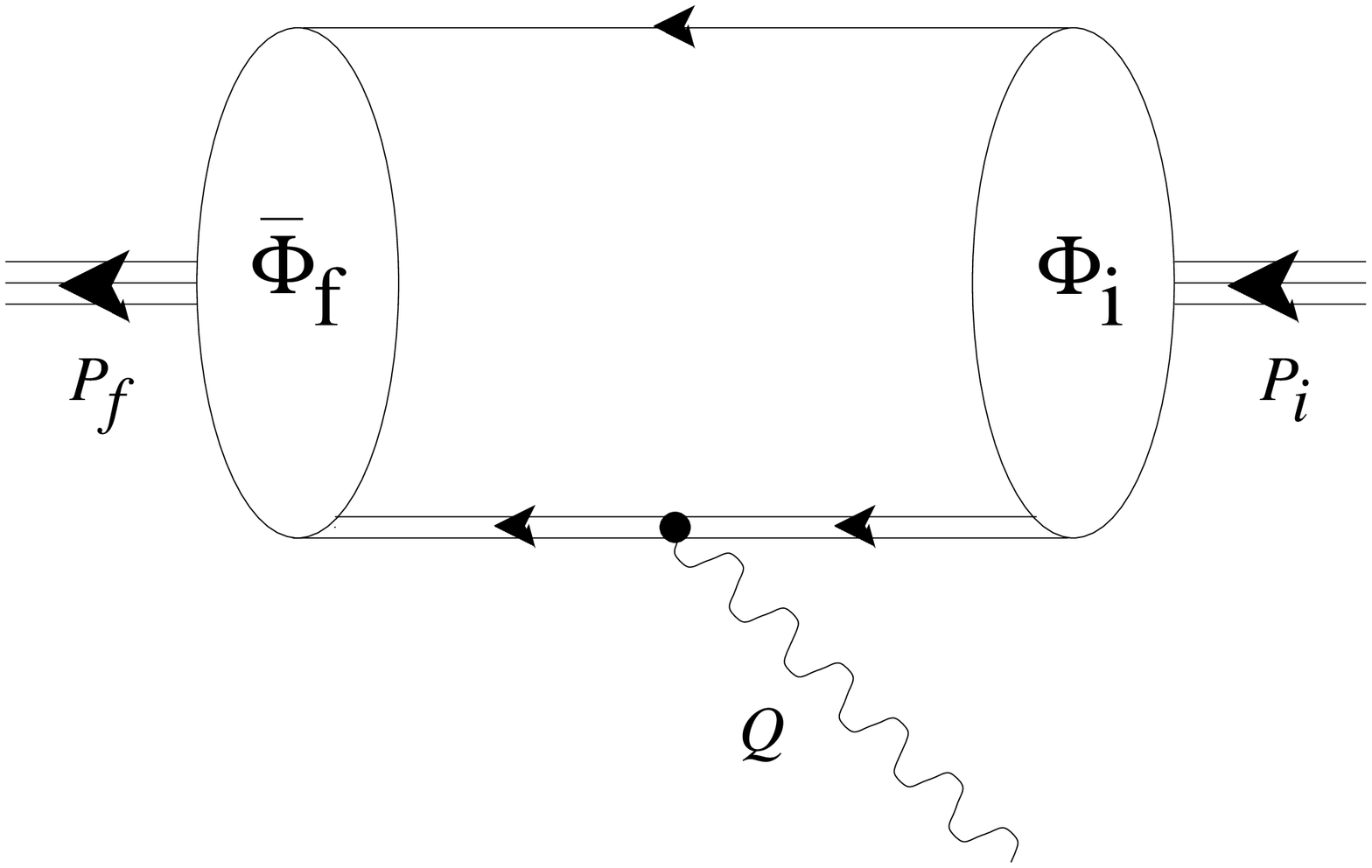,width=4cm}
 \end{minipage}
\end{center}
\caption{Impulse approximation diagrams contributing to the electromagnetic
current.}
\label{impulse}
\end{figure}

The current operator $J^\mu$ in eq.\ (\ref{gdef}) contains
all possible couplings of the photon to the kernel $G$ of the BS equation.
The two simplest contributions make up the impulse approximation
pictorially shown in Fig.\ \ref{impulse}. Since the impulse approximation
does not mix scalar and axialvector amplitudes there are altogether
four diagrams to compute: quark and diquark coupling and for each of the
two the matrix element between scalar and axialvector amplitudes of
the nucleon. Flavor algebra \cite{Kei96b} yields the following
current matrix elements for proton and neutron,
\begin{eqnarray}
 \langle J^{\mu,\text{imp}}_{proton}\rangle & = & \frac{2}{3} \langle J^\mu_{q,s} \rangle +
       \frac{1}{3} \langle J^\mu_{dq,s} \rangle +
                  \langle J^\mu_{dq,a} \rangle, \\
 \langle J^{\mu,\text{imp}}_{neutron}\rangle & = & -\frac{1}{3} \langle J^\mu_{q,s} \rangle +
       \frac{1}{3} \langle J^\mu_{dq,s} \rangle +
      \frac{1}{3} \langle J^\mu_{q,a} \rangle -
      \frac{1}{3} \langle J^\mu_{dq,a} \rangle. 
\end{eqnarray}

\begin{figure}[b]
\begin{center}
 \epsfig{file=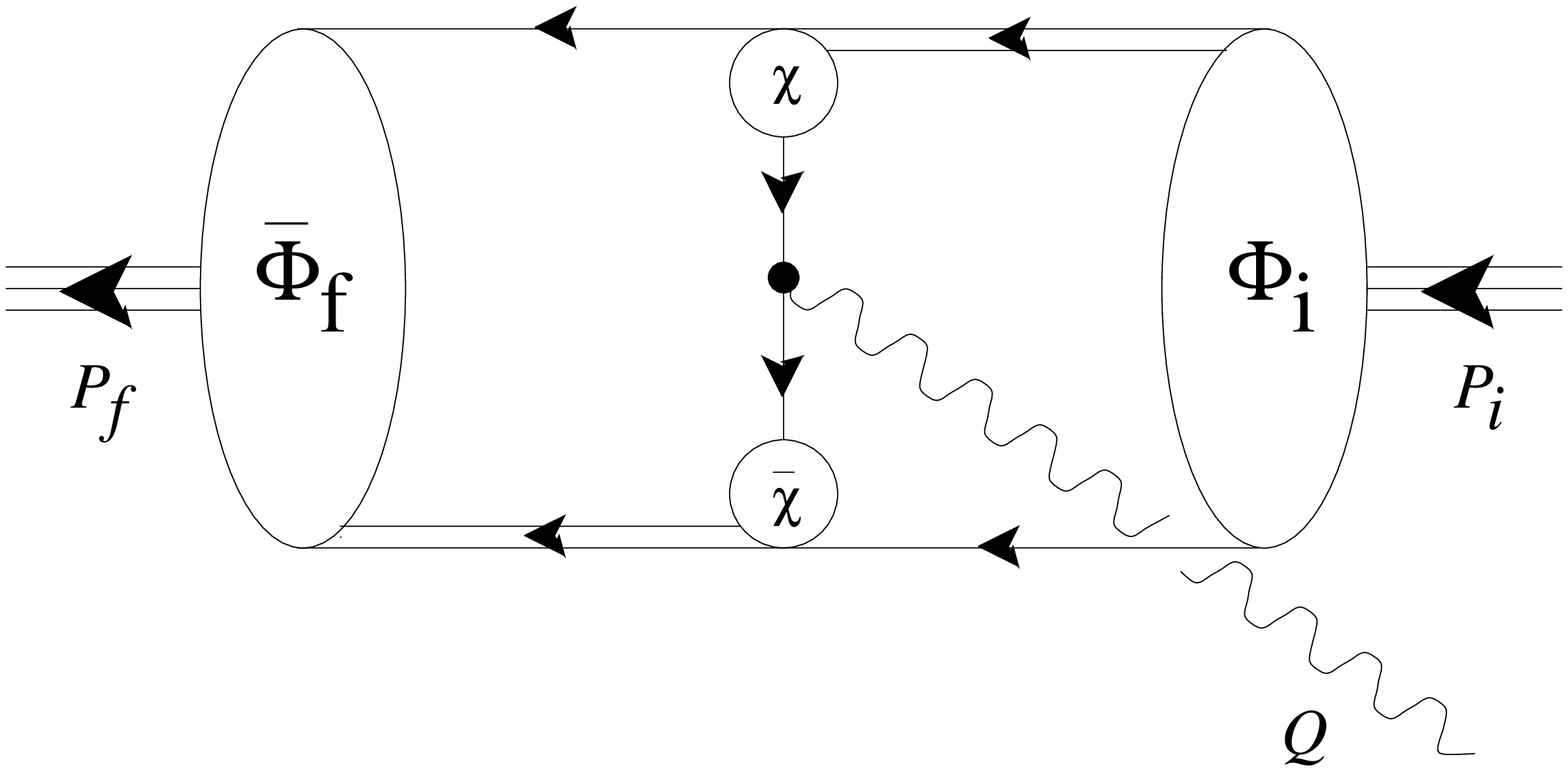,width=4cm}
 \epsfig{file=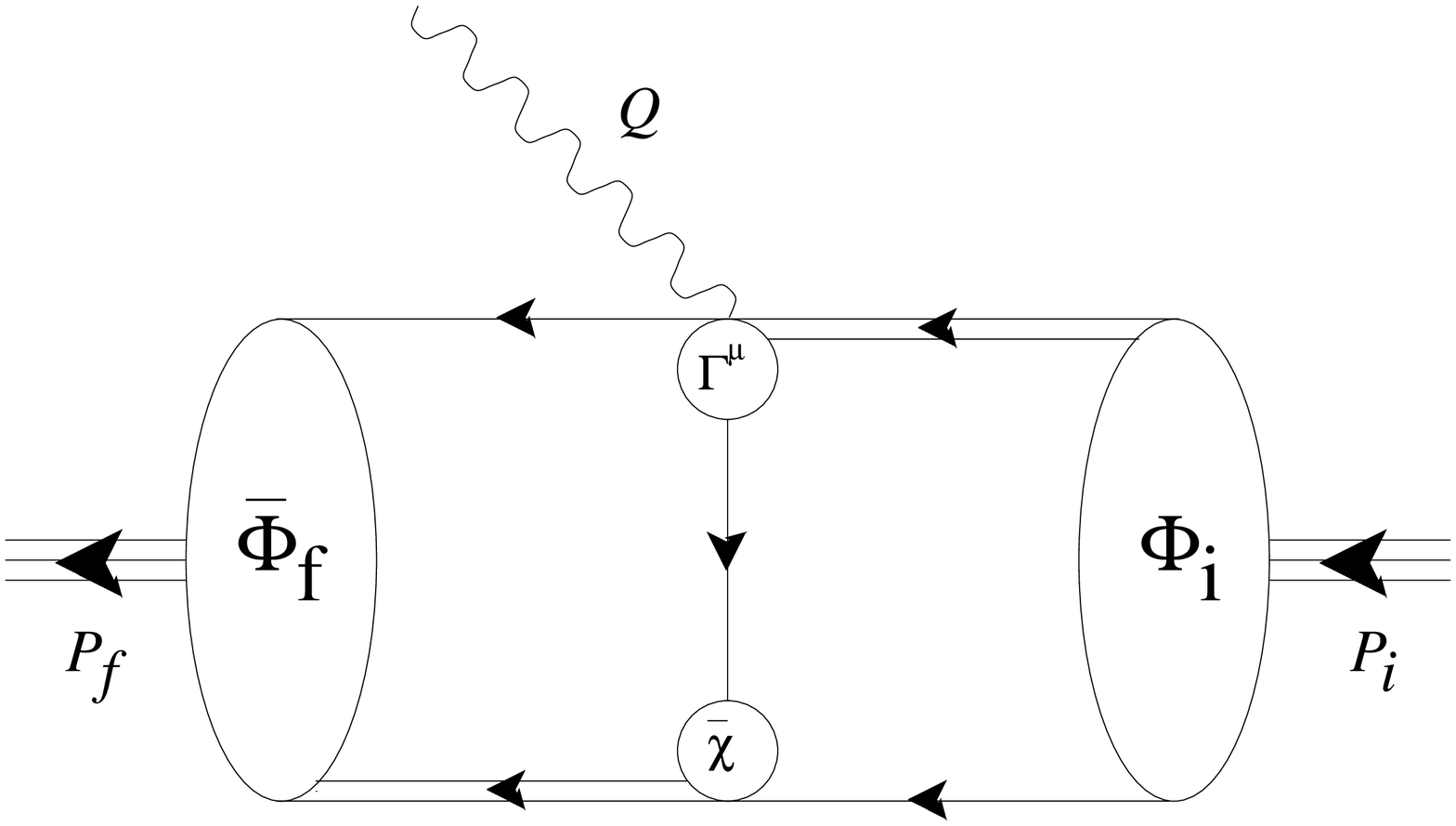,width=4cm}
 \epsfig{file=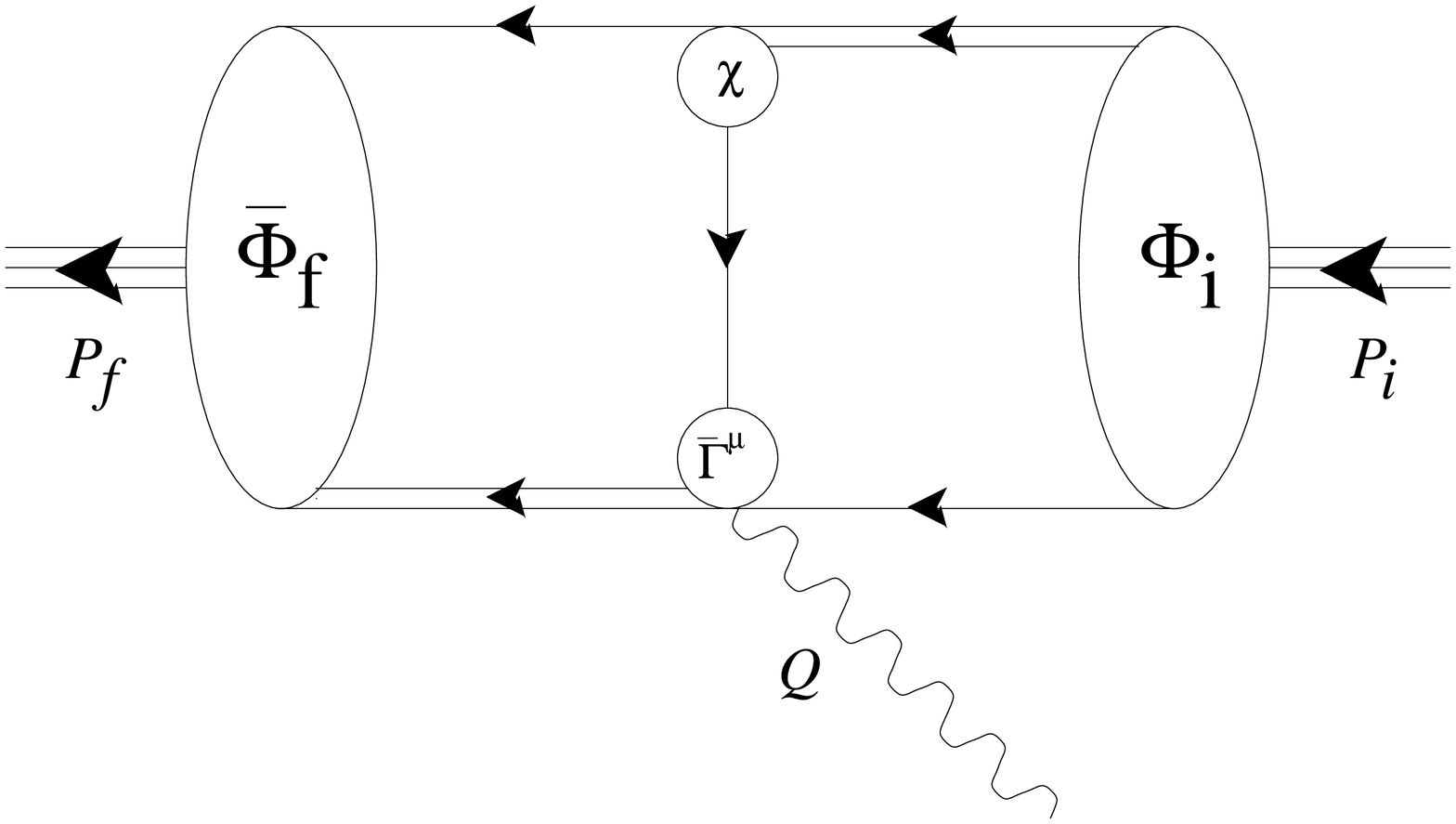,width=4cm}
\end{center}
\caption{Exchange diagrams contributing to the electromagnetic current.}
\label{ex}
\end{figure}

The diagrams of the impulse approximation separately conserve the
current in the Salpeter approximation. The proof depends on 
the behaviour of the current under time reversal and parity transformation
and can be found in \cite{Kei96a} and more explicitly in \cite{Kei97}.
Furthermore there
is the peculiar identity at zero photon momentum transfer,
\begin{equation}
 \langle J^\mu_{q,s[a]} (Q^2=0) \rangle = 
    \langle J^\mu_{dq,s[a]} (Q^2=0)\rangle,
\end{equation}
which guarantees that proton and neutron have their correct charge (i.e.\
$G_{E}^p (Q^2=0)=1$,  $G_{E}^n (Q^2$=0)=0). This is
{\em not} so in the full calculation, to ensure current conservation
and the physical charges one has to take into account all diagrams of
the photon coupling to the interaction part of $G$. These diagrams
are shown in fig. \ref{ex}, the proof of this assertion and the
construction of the ''seagull'' diagrams (photon coupling to the
diquark-quark vertices) can be found in \cite{Oet99}, see also
\cite{Bla99} for a general discussion of the current operator in
three--body theory.

\begin{figure}[t]
 \begin{center}
  \epsfig{file=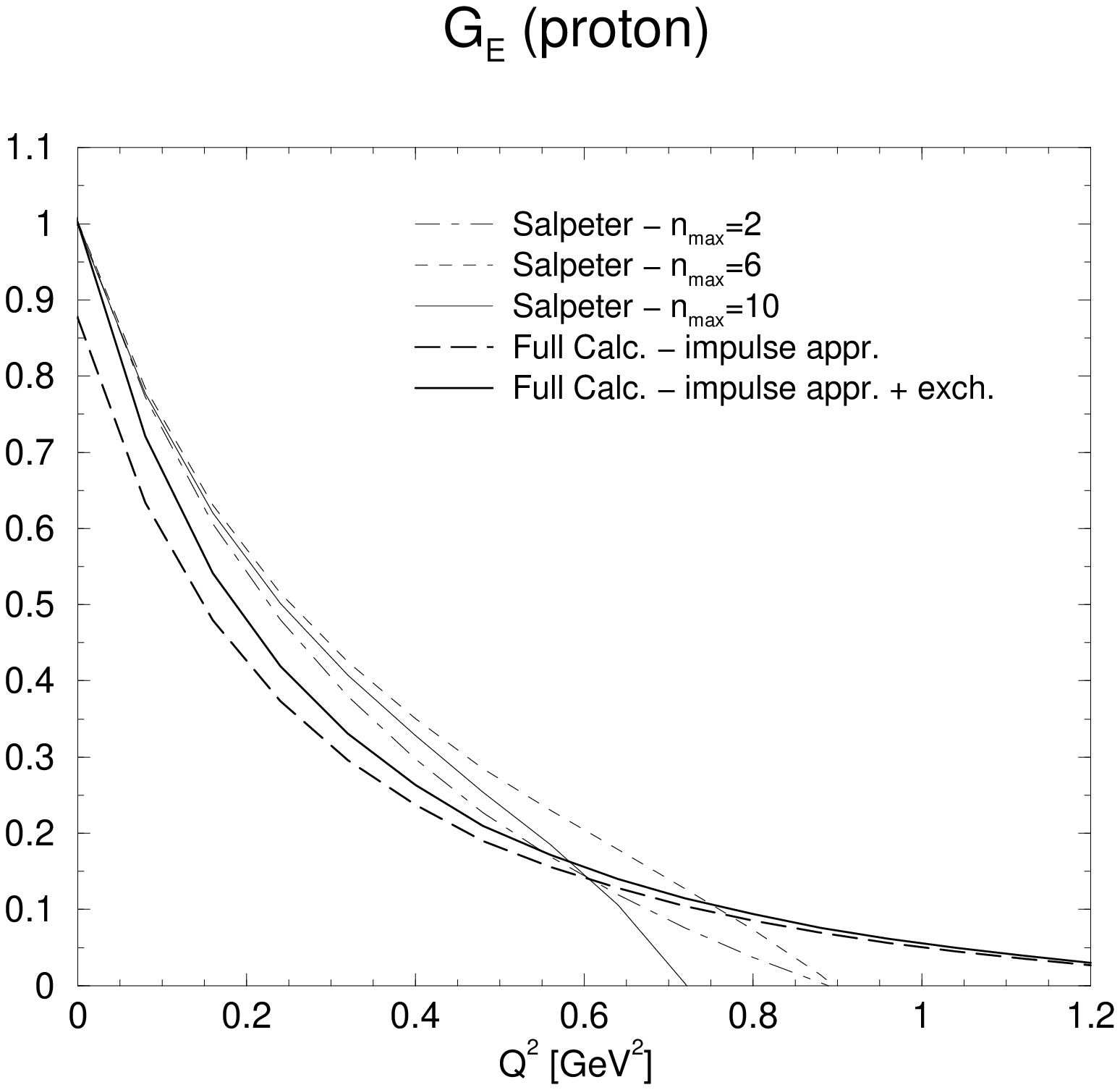,width=6cm}
  \epsfig{file=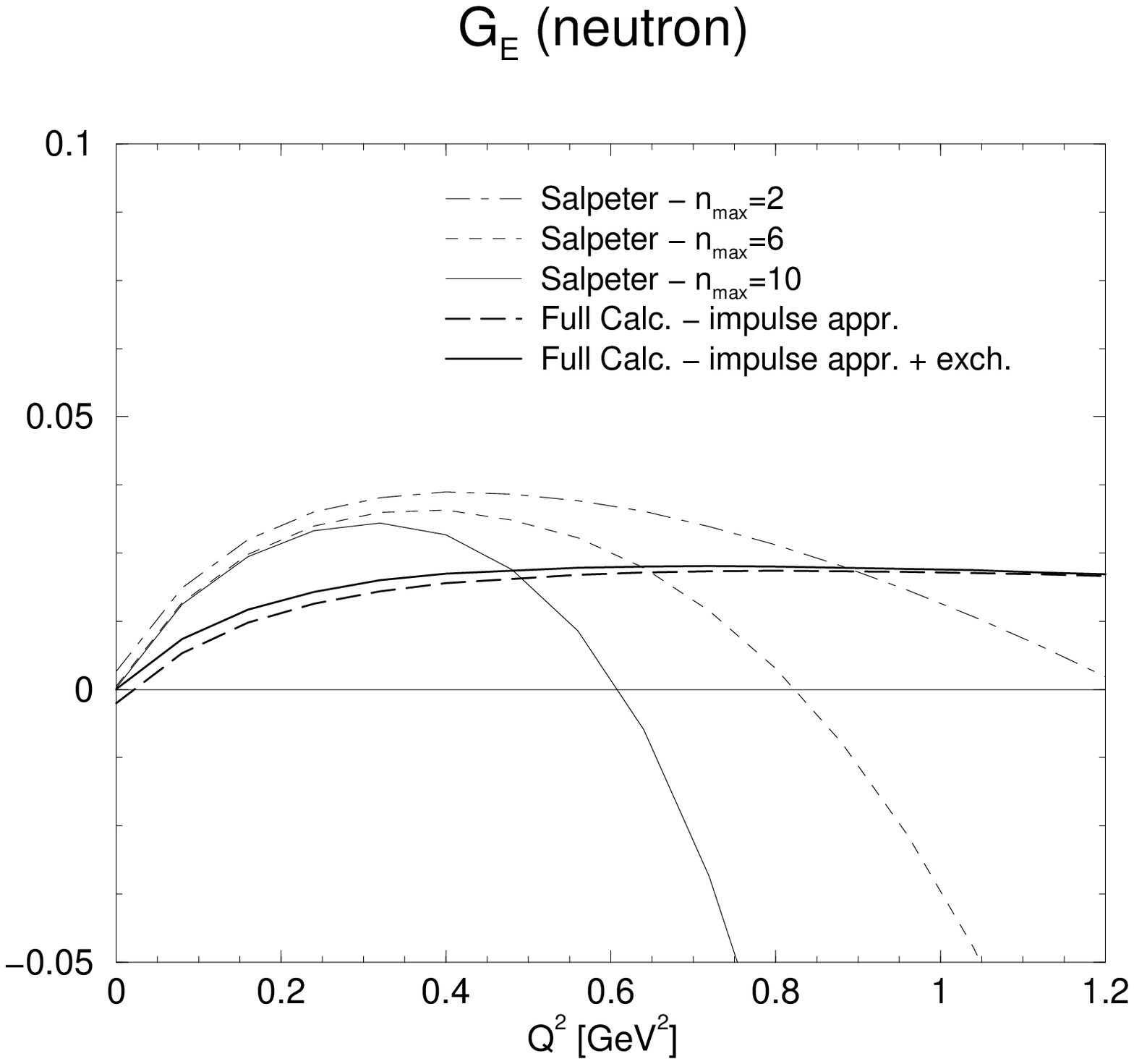,width=6cm}
 \end{center} 
 \caption{Nucleon electric form factors in the Salpeter approximation
and in the full calculation. The curves for the Salpeter approximation
have been obtained by including the Chebyshev moments of the vertex function
up to order $n_{max}$.} 
 \label{ge}
\end{figure}

In fig.\ \ref{ge} the electric form factors of proton and neutron are
displayed, using parameter set II.  The first observation is that in the
Salpeter approximation we could not obtain convergence with the expansion in
Chebyshev polynomials beyond $Q^2 \approx 0.4$ GeV$^2$. We computed the form
factors in the Breit frame where $Q$ is real but $z_i=\hat p_i \cdot \hat P_i$
and  $z_f=\hat p_f \cdot \hat P_f$ have imaginary parts  and their absolute
values may exceed one, except for the case of no momentum transfer
\cite{Oet98}. So this expansion that works in the rest frame, and it does
barely so for the Salpeter approximation,  will not generally work in a moving
frame.  On the other hand, the decrease  of the higher Chebyshev moments is so
drastic for the full four--dimensional solution that form factor calculations
converge easily up to several GeV$^2$ \cite{Oet99,Blo99}. However, as can be
seen from fig. \ref{ge} the Salpeter approximation badly fails above 0.5
GeV$^2$ thereby revealing its semi--relativistic nature.

The second finding concerns the electromagnetic radii. The
Salpeter approximation tends to underestimate the proton charge radius
and to overestimate the absolute value of the neutron charge radius,
see the first two rows of table \ref{static}.
The axialvectors tend to suppress the neutron electric form factor
much more  in the full calculation than in the Salpeter approximation.

\begin{table}[b]
\begin{tabular}{cl|lllll|l}\hline
  Set &  & $\mu_{sc,q}$ & $\mu_{sc,dq}$ & $\mu_{ax,q}$& $\mu_{ax,dq}$ &
       $\mu_{ex}$ & SUM \\ \hline
 I & Salpeter~~~~~   & 1.57 & 0.01 & - & -    & - & 1.58 (1.58)\\
   & full       & 2.04 & 0.01 & - & -    & 0.33& 2.38 \\
  II & Salpeter & 0.92 (1.08) & 0.00 & 0 & 1.37 (1.7) & - & 2.29 (2.78) \\
     & full     & 1.09 & 0.00 & 0 & 0.98 & 0.38~~~~ & 2.45 \\\hline 
\end{tabular}
\caption{Contributions to the proton magnetic moment from the single diagrams.
The first index refers to the nucleon amplitudes involved and the second
to the photon coupling to either quark or diquark. In parentheses
the values of refs.\ \cite{Kei96a,Kei96b} are given.}
\label{magmom}
\end{table} 

Turning to the magnetic moments, the contributions of the various diagrams
are tabulated in table \ref{magmom} for the proton. 
Following \cite{Kei96b} we ascribed to the
axialvector diquark a rather large anomalous magnetic moment of $\kappa$=1.6
which was needed by the cited author to fit the proton magnetic moment.
As already observed earlier, we could reproduce the magnetic moment
for Set I, however, for Set II, the values differ and especially the
coupling to the axialvector diquark is much weaker in our Salpeter
calculation. Rather more interesting is the comparison in the full
calculation between Set I and Set II: The axialvector diquark
improves  the magnetic moment only marginally! The Salpeter approximation 
tends to overestimate this contribution quite drastically.

\begin{table}
\begin{tabular}{cll|cccccccccc}\hline
 Set &&Dia-& $(r_p)_{\rm e}$ & $(r^2_n)_{\rm e}$ & 
        $\mu_p$ & $(r_p)_{\rm m}$ & $\mu_n$ & $(r_n)_{\rm m}$ & 
        $g_{\SSc \pi NN}$ & $r_{\SSc \pi NN}$ & $g_A$ & $r_A$ \\ 
     &&grams& [fm] & [fm$^2$] & & [fm] & & [fm] & & [fm] & & [fm] \\ \hline
 I & Sal. & imp.   
      & 0.89 & -0.28 & 1.58 & 1.04 & -0.77 & 1.05 & 8.96 & 1.04 & 0.93 & 0.93 \\
   & full & imp.       
      & 1.00 & -0.21 & 2.05 & 1.09 & -1.02 & 1.09 &11.71 & 1.07 & 1.16 & 0.99 \\ 
  &    & all 
      & 0.99 & -0.23 & 2.38 & 1.06 & -1.65 & 1.00 &15.34 & 1.04 & 1.46 & 1.02
 \\ \hline
  II & Sal.& imp. 
      & 0.88 & -0.06 & 2.29 & 0.85 & -0.98 & 0.94 & 6.03 & 1.15 & 0.53 & 1.08 \\
     & full&imp.  
      & 1.01 & -0.04 & 2.07 & 0.99 & -1.02 & 1.06 & 6.95 & 1.21 & 0.65 & 1.14 \\
    &  & all
      & 1.01 & -0.04 & 2.45 & 0.99 & -1.26 & 1.05 & 9.36 & 1.15 & 0.82 & 1.10 \\
     \hline 
\end{tabular}
\caption{Some static quantities of the nucleon. The numbers for the Salpeter
 calculation and the ones in the first line for the full calculation are
 obtained with impulse approximation diagrams only. The second line for the
 full calculation includes the couplings to the quark-diquark interaction
 kernel. For $g_a$ and $g_{\pi NN}$, only diagrams with quark couplings 
 have been considered. }
\label{static}
\end{table}
 
Finally we want to mention that the Salpeter approximation underestimates 
the pion--nucleon coupling $g_{\pi NN}$ and the axial coupling $g_A$ quite 
sizeably, see table \ref{static}.
 
\paragraph{Conclusions\\}

In this letter we have presented results for a covariant diquark--quark model.
The ladder BS equation for the nucleon has been solved in a fully
covariant way and in the instantaneous Salpeter approximation. 
As for the model with scalar
diquarks only we have verified the results of ref.\ \cite{Kei96a} whereas there
are discrepancies if the axialvector diquark is included. Part of these
differences are due to the fact that in ref.\ \cite{Kei96b} not all 
(ground state) axialvector components have been taken into account.
Additionally, we take our result as an indication that the calculations
presented in ref.\ \cite{Kei96b} might suffer from some minor error.

The main purpose of this letter is the comparison of observables calculated in
the Salpeter approximation to the ones obtained in the fully four--dimensional
scheme. The first very surprising observation is the overestimation of the BS
eigenvalue in the Salpeter approximation. Phrased otherwise, for a given
coupling constant the binding energy would be much too small in the Salpeter 
approximation.  We have also demonstrated that the Salpeter approximation
violates badly the approximate $O$(4) symmetry of the BS equation. This has
drastic consequences for the resulting nucleon electromagnetic form factors if
the  photon virtuality exceeds 0.4 GeV$^2$. Whereas different nucleon radii
differ  only mildly in these two approaches one sees very clearly that the
results for the magnetic moments, the pion--nucleon coupling and the axial
coupling are underestimated in the Salpeter approximation.

\paragraph{Acknowledgement\\}

We thank S.\ Ahlig, H.\ Reinhardt and H.\ Weigel 
for a critical reading of the manuscript and their comments.

\end{document}